# Fisher, Neyman-Pearson or NHST? A Tutorial for Teaching Data Testing

**Jose D. Perezgonzalez**[*]

Massey University, New Zealand

**\* Correspondence:**
Dr. Jose D. Perezgonzalez
Massey University,
Business School, SST8.18,
P.O.Box 11-222, Palmerston North 4442, New Zealand.
j.d.perezgonzalez@massey.ac.nz

## Abstract

Despite frequent calls for the overhaul of null hypothesis significance testing (NHST), this controversial procedure remains ubiquitous in behavioral, social and biomedical teaching and research. Little change seems possible once the procedure becomes well ingrained in the minds and current practice of researchers; thus, the optimal opportunity for such change is at the time the procedure is taught, be this at undergraduate or at postgraduate levels. This paper presents a tutorial for the teaching of data testing procedures, often referred to as hypothesis testing theories. The first procedure introduced is Fisher's approach to data testing—tests of significance; the second is Neyman-Pearson's approach—tests of acceptance; the final procedure is the incongruent combination of the previous two theories into the current approach—NSHT. For those researchers sticking with the latter, two compromise solutions on how to improve NHST conclude the tutorial.

## 1. Introduction

This paper introduces the classic approaches for testing research data: tests of significance, which Fisher helped develop and promote starting in 1925; tests of statistical hypotheses, developed by Neyman and Pearson in 1928; and null hypothesis significance testing (NHST), first concocted by Lindquist in 1940. This chronological arrangement is fortuitous insofar it introduces the simpler testing approach by Fisher first, then moves onto the more complex one by Neyman and Pearson, before tackling the incongruent hybrid approach represented by NHST (Gigerenzer, 2004; Hubbard, 2004). Other theories, such as Bayes's hypotheses testing (Lindley, 1965) and Wald's (1950) decision theory, are not object of this tutorial.

The main aim of the tutorial is to illustrate the bases of discord in the debate against NHST (Gigerenzer, 2004; Macdonald, 2002), which remains a problem not only yet unresolved but very much ubiquitous in current data testing (e.g., Franco, Malhotra, and Simonovits, 2014) and teaching (e.g., Dancey and Reidy, 2014), especially in the biological sciences (Lovel, 2013; Ludbrook, 2013), social sciences (Frick, 1996), psychology (Gigerenzer, 2004; Nickerson, 2000) and education (Carver, 1978, 1993).

This tutorial is appropriate for the teaching of data testing at undergraduate and postgraduate levels, and is best introduced when students are knowledgeable on important background information regarding research methods (such as random sampling) and inferential statistics (such as frequency distributions of means).

In order to improve understanding, statistical constructs that may bring about confusion between theories are labelled differently, attending to their function in preference to their historical use (Perezgonzalez, 2014). Descriptive notes (notes) and caution notes (caution) are provided to clarify matters whenever appropriate.

## 2.      Fisher's approach to data testing

Ronald Aylmer Fisher was the main force behind tests of significance (Neyman, 1967) and can be considered the most influential figure in the current approach to testing research data (Hubbard, 2004). Although some steps in Fisher's approach may be worked out a priori (e.g., the setting of hypotheses and levels of significance), the approach is eminently inferential and all steps can be set up a posteriori, once the research data are ready to be analyzed (Fisher, 1955; Macdonald, 1997). Some of these steps can even be omitted in practice, as it is relatively easy for a reader to recreate them. Fisher's approach to data testing can be summarized in the five steps described below.

**Step 1 - Select an appropriate test.** This step calls for selecting a test appropriate to, primarily, the research goal of interest (Fisher, 1932), although you may also need to consider other issues, such as the way your variables have been measured. For example, if your research goal is to assess differences in the number of people in two independent groups, you would choose a chi-square test (it requires variables measured at nominal levels); on the other hand, if your interest is to assess differences in the scores that the people in those two groups have reported on a questionnaire, you would choose a *t*-test (it requires variables measured at interval or ratio levels and a close-to-normal distribution of the groups' differences).

**Step 2 - Set up the null hypothesis ($H_0$).** The null hypothesis derives naturally from the test selected in the form of an exact statistical hypothesis (e.g., $H_0$: $M_1 - M_2 = 0$; Carver, 1978; Frick, 1996; Neyman and Pearson, 1933). Some parameters of this hypothesis, such as variance and degrees of freedom, are estimated from the sample, while other parameters, such as the distribution of frequencies under a particular distribution, are deduced theoretically. The statistical distribution so established thus represents the random variability that is theoretically expected for a statistical nil hypothesis (i.e., $H_0 = 0$) given a particular research sample (Bakan, 1966; Fisher, 1954, 1955; Hubbard, 2004; Macdonald, 2002). It is called the null hypothesis because it stands to be nullified with research data (Gigerenzer, 2004).

Among things to consider when setting the null hypothesis is its directionality.

**Directional and non-directional hypotheses.** With some research projects, the direction of the results is expected (e.g., one group will perform better than the other). In these cases, a directional null hypothesis covering all remaining possible results can be set (e.g., $H_0$: $M_1 - M_2 \leq 0$). With other projects, however, the direction of the results is not predictable or of no research interest. In these cases, a non-directional hypothesis is most suitable (e.g., $H_0$: $M_1 - M_2 = 0$).

*Notes: $H_0$ does not need to be a nil hypothesis, that is, one that always equals zero (Fisher, 1955; Gigerenzer, 2004). For example, $H_0$ could be that the group difference is not larger than certain value (Newman et al., 2001). More often than not, however, $H_0$ tends to be zero.*

*Setting up $H_0$ is one of the steps usually omitted if following the typical nil expectation (e.g., no correlation between variables, no differences in variance among groups, etc). Even directional nil hypotheses are often omitted, instead specifying that one-tailed tests (see below) have been used in the analysis.*

**Step 3 - Calculate the probability of the results under $H_0$ ($p$).** Once the corresponding theoretical distribution is established, the probability ($p$-value) of any datum under the null hypothesis is also established, which is what statistics calculate (Bakan, 1966; Cortina and Dunlap, 1997; Fisher, 1955, 1960; Hagen, 1997; Johnstone, 1987). Data closer to the mean of the distribution (Figure 1) have a greater probability of occurrence under the null distribution; that is, they appear more frequently and show a larger $p$-value (e.g., p = .46, or 46 times in a hundred trials). On the other hand, data located further away from the mean have a lower probability of occurrence under the null distribution; that is, they appear less often and, thus, show a smaller $p$-value (e.g., p = .003). Of interest to us is the probability of our research results under such null distribution (e.g., the probability of the difference in means between two research groups).

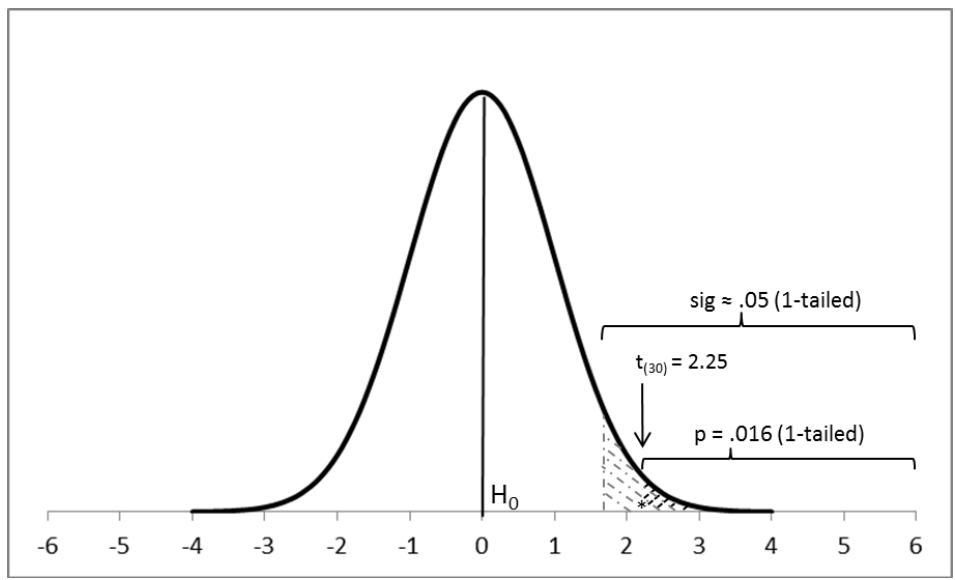

*Figure 1*. Location of a $t$ value and its corresponding $p$-value on a theoretical $t$ distribution with 30 degrees of freedom. The actual $p$-value conveys stronger evidence against $H_0$ than sig ≈ .05 and can be considered highly significant.

The $p$-value comprises the probability of the observed results and also of any other more extreme results (e.g., the probability of the actual difference between groups and any other difference more extreme than that). Thus, the $p$-value is a cumulative probability rather than an exact point probability: It covers the probability area extending from the observed results towards the tail of the distribution (Carver, 1978; Fisher, 1960; Frick, 1996; Hubbard, 2004).

> *Note:* P-*values provide information about the theoretical probability of the observed and more extreme results under a null hypothesis assumed to be true (Bakan, 1966; Fisher, 1960), or, said otherwise, the probability of the data given a true hypothesis—P(D/H); Carver, 1978;* Hubbard, 2004. *As $H_0$ is always true (i.e., it shows the theoretical random distribution of frequencies under certain parameters), it cannot, at the same time, be false nor falsifiable a posteriori. Basically, if at any point you say that $H_0$ is false, then you are also invalidating the whole test and its results. Furthermore, because $H_0$ is always true, it cannot be proved, either.*

**Step 4 - Assess the statistical significance of the results.** Fisher proposed tests of significance as a tool for identifying research results of interest, defined as those with a low probability of occurring as mere random variation of a null hypothesis. A research result with a low *p*-value may, thus, be taken as evidence against the null (i.e., as evidence that it may not explain those results satisfactorily; Bakan, 1966; Fisher, 1960; Johnstone, 1987; Macdonald, 2002). How small a result ought to be in order to be considered statistically significant is largely dependent on the researcher in question, and may vary from research to research (Fisher, 1960; Gigerenzer, 2004). The decision can also be left to the reader, so reporting exact *p*-values is very informative (Fisher, 1973; Gigerenzer, 2004; Macdonald, 1997).

Overall, however, the assessment of research results is largely made bound to a given level of significance, by comparing whether the research *p*-value is smaller than such level of significance or not (Fisher, 1954, 1960; Johnstone, 1987):

- If the *p*-value is approximately equal to or smaller than the level of significance, the result is considered statistically significant.

- If the *p*-value is larger than the level of significance, the result is considered statistically non-significant.

Among things to consider when assessing the statistical significance of research results are the level of significance, and how it is affected by the directionality of the test and other corrections.

**Level of significance (sig).** The level of significance is a theoretical *p*-value used as a point of reference to help identify statistically significant results (Figure 1). There is no need to set up a level of significance a priori nor for a particular level of significance to be used in all occasions, although levels of significance such as 5% (sig ≈ .05) or 1% (sig ≈ .01) may be used for convenience, especially with novel research projects (Carver, 1978; Fisher, 1960; Gigerenzer, 2004). This highlights an important property of Fisher's levels of significance: They do not need to be rigid (e.g., *p*-values such as .049 and .051 have about the same statistical significance around a convenient level of significance of 5%; Johnstone, 1987).

Another property of tests of significance is that the observed *p*-value is taken as evidence against the null hypothesis, so that the smaller the *p*-value the stronger the evidence it provides (Fisher, 1960; Spielman, 1978). This means that it is plausible to gradate the strength of such evidence with smaller levels of significance. For example, if using 5% (sig ≈ .05) as a convenient level for identifying results which are just significant, then 1% (sig ≈ .01) may be used as a convenient level for identifying highly significant results and 1‰ (sig ≈ .001) for identifying extremely significant results.

> **Notes:** *Setting up a level of significance is another step usually omitted. In such cases, you may assume the researcher is using conventional levels of significance.*
>
> *If both $H_0$ and sig are made explicit, they could be joined in a single postulate, such as $H_0$: $M_1 - M_2 = 0$, sig $\approx .05$.*
>
> *Notice that the* p-value *informs about the probability associated with a given test value (e.g., a* t *value). You could use this test value to decide about the significance of your results in a fashion similar to Neyman-Pearson's approach (see below). However, you get more information about the strength of the research evidence with* p-values.
>
> *Although the* p-value *is the most informative statistic of a test of significance, in psychology (e.g., American Psychological Association, 2010) you also report the research value of the test—e.g. t(30) = 2.25, p = .016, 1-tailed. Albeit cumbersome and largely ignored by the reader, the research value of the test offers potentially useful information (e.g., about the valid sample size used with a test).*
>
> **Caution:** *Be careful not to interpret Fisher's* p-values *as Neyman-Pearson's Type I errors (α, see below). Probability values in single research projects are not the same than probability values in the long run (Johnstone, 1987), something illustrated by Berger (2003)—who reported that p = .05 often corresponds to α = .5 (or anywhere between α = .22 and α > .5)—and Cumming (2014)—who simulates the 'dance' of p-values in the long run, commented further in Perezgonzalez, 2015).*

**One-tailed and two-tailed tests.** With some tests (e.g., *F*-tests) research data can only be tested against one side of the null distribution (one-tailed tests), while other tests (e.g., *t*-tests) can test research data against both sides of the null distribution at the same time. With one-tailed tests you set the level of significance on the appropriate tail of the distribution. With two-tailed tests you cover both eventualities by dividing the level of significance between both tails (Fisher, 1960; Macdonald, 1997), which is commonly done by halving the total level of significance in two equal areas (thus covering, for example, the 2.5% most extreme positive differences and the 2.5% most extreme negative differences).

> **Note:** *The tail of a test depends on the test in question, not on whether the null hypothesis is directional or non-directional. However, you can use two-tailed tests as one-tailed ones when testing data against directional hypotheses.*

**Correction of the level of significance for multiple tests.** As we introduced earlier, a *p*-value can be interpreted in terms of its expected frequency of occurrence under the specific null distribution for a particular test (e.g., p = .02 describes a result that is expected to appear 2 times out of 100 under $H_0$). The same goes for theoretical *p*-values used as levels of significance. Thus, if more than one test is performed, this has the consequence of also increasing the probability of finding statistical significant results which are due to mere chance variation. In order to keep such probability at acceptable levels overall, the level of significance may be corrected downwards (Hagen, 1997). A popular correction is Bonferroni's, which reduces the level of significance proportionally to the number of tests carried out. For example, if your selected level of significance is 5% (sig $\approx$ .05) and you carry out two tests, then such level of significance is maintained overall by correcting the level of significance for each test down to 2.5% (sig $\approx$ .05 / 2 tests $\approx$ .025, or 2.5% per test).

> *Note: Bonferroni's correction is popular but controversial, mainly because it is too conservative, more so as the number of multiple tests increases. There are other methods for controlling the probability of false results when doing multiple comparisons, including familywise error rate methods (e.g. Holland and Copenhaver, 1987), false discovery rate methods (e.g., Benjamini and Hochberg, 1995), resampling methods (jackknifing, bootstrapping— e.g., Efron, 1981), and permutation tests (i.e., exact tests—e.g., Gill, 2007).*

**Step 5 - Interpret the statistical significance of the results.** A significant result is literally interpreted as a dual statement: Either a rare result that occurs only with probability $p$ (or lower) just happened, or the null hypothesis does not explain the research results satisfactorily (Carver, 1978; Fisher, 1956; Johnstone, 1987; Macdonald, 1997). Such literal interpretation is rarely encountered, however, and most common interpretations are in the line of 'The null hypothesis did not seem to explain the research results well, thus we inferred that other processes—which we believe to be our experimental manipulation—exist that account for the results', or 'The research results were statistically significant, thus we inferred that the treatment used accounted for such difference'.

Non-significant results may be ignored (Fisher, 1960; Nunnally, 1960), although they can still provide useful information, such as whether results were in the expected direction and about their magnitude (Fisher, 1955). In fact, although always denying that the null hypothesis could ever be supported or established, Fisher conceded that non-significant results might be used for confirming or strengthening it (Fisher, 1955; Johnstone, 1987).

> *Note: Statistically speaking, Fisher's approach only ascertains the probability of the research data under a null hypothesis. Doubting or denying such hypothesis given a low* p*-value does not necessarily "support" or "prove" that the opposite is true (e.g., that there is a difference or a correlation in the population). More importantly, it does not "support" or "prove" that whatever else has been done in the research (e.g., the treatment used) explains the results, either (Macdonald, 1997). For Fisher, a good control of the research design (Cortina and Dunlap, 1997; Fisher, 1955; Johnstone, 1987), especially random allocation, is paramount to make sensible inferences based on the results of tests of significance (Fisher, 1954; Neyman, 1967). He was also adamant that, given a significant result, further research was needed to establish that there has indeed been an effect due to the treatment used (Fisher, 1954; Johnstone, 1987; Macdonald, 2002). Finally, he considered significant results as mere data points and encouraged the use of meta-analysis for progressing further, combining significant and non-significant results from related research projects (Fisher, 1960; Neyman, 1967).*

## 2.1.    Highlights of Fisher's approach

**Flexibility.** Because most of the work is done a posteriori, Fisher's approach is quite flexible, allowing for any number of tests to be carried out and, therefore, any number of null hypotheses to be tested (a correction of the level of significance may be appropriate, though—Macdonald, 1997).

**Better suited for ad-hoc research projects.** Given above flexibility, Fisher's approach is well suited for single, ad-hoc, research projects (Johnstone, 1987; Neyman, 1956), as well as for exploratory research (Frick, 1996; Gigerenzer, 2004; Macdonald, 1997).

**Inferential.** Fisher's procedure is largely inferential, from the sample to the population of reference, albeit of limited reach, mainly restricted to populations that share parameters similar to those estimated from the sample (Fisher, 1954, 1955; Hubbard, 2004; Macdonald, 2002).

**No power analysis.** Neyman (1967) and Kruskal and Savage (Kruskal, 1980) were surprised that Fisher did not explicitly attend to the power of a test. Fisher talked about sensitiveness, a similar concept, and how it could be increased by increasing sample size (Fisher, 1960). However, he never created a mathematical procedure for controlling sensitiveness in a predictable manner (Hubbard, 2004; Macdonald, 1997).

**No alternative hypothesis.** One of the main critiques to Fisher's approach is the lack of an explicit alternative hypothesis (Gigerenzer, 2004; Hubbard, 2004; Macdonald, 2002), because there is no point in rejecting a null hypothesis without an alternative explanation being available (Pearson, 1990). However, Fisher considered alternative hypotheses implicitly—these being the negation of the null hypotheses—so much so that for him the main task of the researcher—and a definition of a research project well done—was to systematically reject with enough evidence the corresponding null hypothesis (Fisher, 1960).

## 3.      Neyman-Pearson's approach to data testing

Jerzy Neyman and Egon Sharpe Pearson tried to improve Fisher's procedure (Fisher, 1955; Jones and Tukey, 2000; Macdonald, 2002; Pearson, 1955) and ended up developing an alternative approach to data testing. Neyman-Pearson's approach is more mathematical than Fisher's and does much of its work a priori, at the planning stage of the research project (Fisher, 1955; Gigerenzer, 2004; Hubbard, 2004; Macdonald, 1997). It also introduces a number of constructs, some of which are similar to those of Fisher. Overall, Neyman-Pearson's approach to data testing can be considered tests of acceptance (Fisher, 1955; Pearson, 1955; Spielman, 1978; Perezgonzalez, 2014), summarized in the following eight main steps.

### 3.1.    A priori steps

**Step 1 - Set up the expected effect size in the population.** The main conceptual innovation of Neyman-Pearson's approach was the consideration of explicit alternative hypotheses when testing research data (Gigerenzer, 2004; Hubbard, 2004; Macdonald, 2002; Neyman, 1956; Neyman and Pearson, 1928, 1933). In their simplest postulate, the alternative hypothesis represents a second population that sits alongside the population of the main hypothesis on the same continuum of values. These two groups differ by some degree: the effect size (Cohen, 1988; Macdonald, 1997).

Although the effect size was a new concept introduced by Neyman and Pearson, in psychology it was popularized by Cohen (1988). For example, Cohen's conventions for capturing differences between groups—d (Figure 2)—were based on the degree of visibility of such differences in the population: the smaller the effect size, the more difficult to appreciate such differences; the larger the effect size, the easier to appreciate such differences. Thus, effect sizes also double as a measure of importance in the real world (Cohen, 1988; Frick, 1996; Nunnally, 1960).

When testing data about samples, however, statistics do not work with unknown population distributions but with distributions of samples, which have narrower standard errors. In these cases, the effect size can still be defined as above because the means of the populations remain unaffected, but the sampling distributions would appear separated rather than overlapping (Figure 3). Because we rarely know the parameters of populations, it is their equivalent effect size measures in the context of sampling distributions which are of interest.

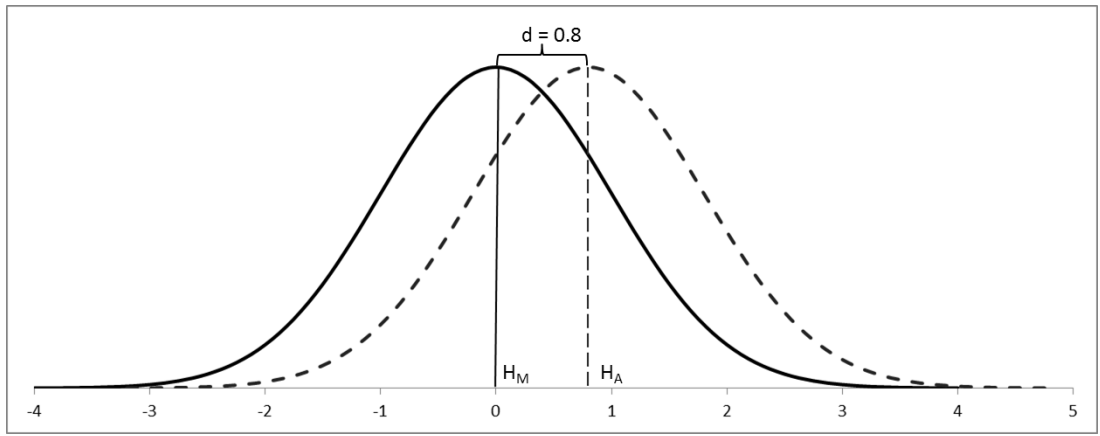

*Figure 2.* A conventional large difference—Cohen's *d = 0.8*—between two normally distributed populations, as a fraction of one standard deviation.

As we shall see below, the alternative hypothesis is the one that provides information about the effect size to be expected. However, because this hypothesis is not tested, Neyman-Pearson's procedure largely ignores its distribution except for a small percentage of it, which is called 'beta' (β; Gigerenzer, 2004). Therefore, it is easier to understand Neyman-Pearson's procedure if we peg the effect size to beta and call it the expected minimum effect size (MES; Figure 3). This helps us conceptualize better how Neyman-Pearson's procedure works (Schmidt, 1996): The minimum effect size effectively represents that part of the main hypothesis that is not to be rejected by the test (i.e., MES captures values of no research interest which you want to left under $H_M$; Cortina and Dunlap, 1997; Hagen, 1997; Macdonald, 2002). (Worry not, as there is no need to perform any further calculations: The population effect size is the one to use, for example, for estimating research power.)

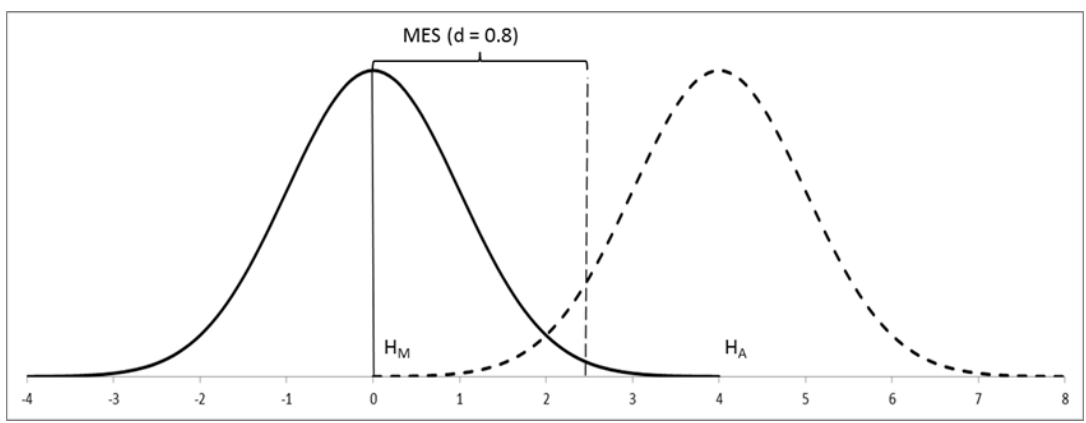

*Figure 3.* Sampling distributions (N = 50 each) of the populations in Figure 2. MES (*d* = 0.8), assuming β = .20, is *d* = 0.32 (i.e., the expected difference in the population ranges between *d* = 0.32 and infinity).

*Note:* A particularity of Neyman-Pearson's approach is that the two hypotheses are assumed to represent defined populations, the research sample being an instance of either of them (i.e., they are populations of samples generated by repetition of a common random process—Hagen, 1997; Hubbard, 2004; Neyman and Pearson, 1928; Pearson, 1955). This is unlike Fisher's population, which can be considered more theoretical, generated ad-hoc so as for providing the appropriate random distribution for the research sample at hand (i.e., a population of samples similar to the research sample—Fisher, 1955; Johnstone, 1987).

**Step 2 - Select an optimal test.** As we shall see below, another of Neyman-Pearson's contributions was the construct of the power of a test. A spin-off of this contribution is that it has been possible to establish which tests are most powerful (for example, parametric tests are more powerful than non-parametric tests, and one-tailed tests are more powerful than two-tailed tests), and under which conditions (for example, increasing sample size increases power). For Neyman and Pearson, thus, you are better off choosing the most powerful test for your research project (Neyman, 1942, 1956).

**Step 3 - Set up the main hypothesis ($H_M$).** Neyman-Pearson's approach considers, at least, two competing hypotheses, although it only tests data under one of them. The hypothesis which is the most important for the research (i.e., the one you do not want to reject too often) is the one tested (Neyman, 1942; Neyman and Pearson, 1928; Spielman, 1973). This hypothesis is better off written so as for incorporating the minimum expected effect size within its postulate (e.g., $H_M$: $M_1 - M_2 = 0 \pm$ MES), so that it is clear that values within such minimum threshold are considered reasonably probable under the main hypothesis, while values outside such minimum threshold are considered as more probable under the alternative hypothesis (Figure 4).

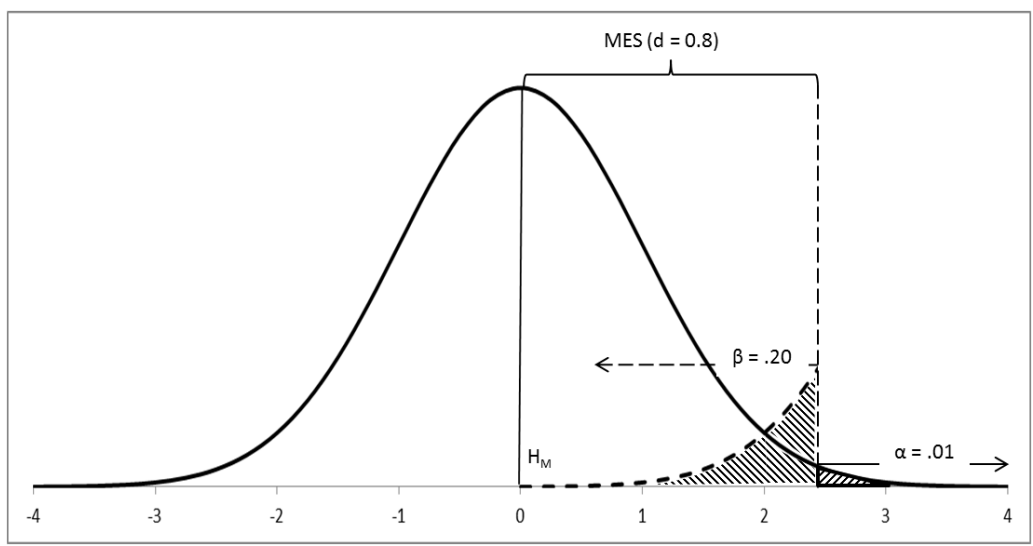

*Figure 4*. Neyman-Pearson's approach tests data under $H_M$ using the rejection region delimited by α. $H_A$ contributes MES and β. Differences of research interest will be equal or larger than MES and will fall within this rejection region.

> *Caution: Neyman-Pearson's $H_M$ is very similar to Fisher's $H_0$. Indeed, Neyman and Pearson also called it the null hypothesis and often postulated it in a similar manner (e.g., as $H_M$: $M_1 - M_2 = 0$). However, this similarity is merely superficial on three accounts: $H_M$ needs to be considered at the design stage ($H_0$ is rarely made explicit); it is implicitly designed to incorporate any value below the MES—i.e., the a priori power analysis of a test aims to capture such minimum difference (effect sizes are not part of Fisher's approach); and it is but one of two competing explanations for the research results ($H_0$ is the only hypothesis, to be nullified with evidence).*

The main aspect to consider when setting the main hypothesis is the Type I error you want to control for during the research.

**Type I error.** A Type I error (or error of the first class) is made every time the main hypothesis is wrongly rejected (thus, every time the alternative hypothesis is wrongly accepted). Because the hypothesis under test is your main hypothesis, this is an error that you want to minimize as much as possible in your lifetime research (Macdonald, 1997; Neyman, 1942; Neyman and Pearson, 1928, 1933).

> *Caution: A Type I error is possible under Fisher's approach, as it is similar to the error made when rejecting $H_0$ (Carver, 1978). However, this similarity is merely superficial on two accounts: Neyman and Pearson considered it an error whose relevance only manifests itself in the long run because it is not possible to know whether such an error has been made in any particular trial (Fisher's approach is eminently ad-hoc, so the risk of a long-run Type I error is of little relevance); therefore, it is an error that needs to be considered and minimized at the design stage of the research project in order to ensure good power—you cannot minimize this error a posteriori (with Fisher's approach, the potential impact of errors on individual projects is better controlled by correcting the level of significance as appropriate, for example, with a Bonferroni correction).*

**Alpha ($\alpha$).** Alpha is the probability of committing a Type I error in the long run (Gigerenzer, 2004). Neyman and Pearson often worked with convenient alpha levels such as 5% ($\alpha = .05$) and 1% ($\alpha = .01$), although different levels can also be set. The main hypothesis can, thus, be written so as for incorporating the alpha level in its postulate (e.g., $H_M$: $M_1 - M_2 = 0 \pm MES$, $\alpha = .05$), to be read as the probability level at which the main hypothesis will be rejected in favor of the alternative hypothesis.

> *Caution: Neyman-Pearson's $\alpha$ looks very similar to Fisher's sig. Indeed, Neyman and Pearson also called it the significance level of the test and used the same conventional cut-off points (5%, 1%). However, this similarity is merely superficial on three accounts: $\alpha$ needs to be set a priori (not necessarily so under Fisher's approach); Neyman-Pearson's approach is not a test of significance (they are not interested in the strength of the evidence against $H_M$) but a test of acceptance (deciding whether to accept $H_A$ instead of $H_M$); and $\alpha$ does not admit gradation—ie, you may choose, for example, either $\alpha = .05$ or $\alpha = .01$, but not both, for the same test (while with Fisher's approach you can have different levels of more extreme significance) .*

**The critical region ($CR_{test}$) and critical value ($CV_{test}$, $Test_{crit}$) of a test.** The alpha level helps draw a critical region, or rejection region (Figure 4), on the probability distribution of the main hypothesis (Neyman and Pearson, 1928). Any research value that falls outside this critical region will be taken as reasonably probable under the main hypothesis, and any research result that falls within the critical region will be taken as most probable under the alternative hypothesis. The alpha level, thus, also helps identify the location of the critical value of such test, the boundary for deciding between hypotheses. Thus, once the critical value is known—see below—, the main hypothesis can also be written so as for incorporating such critical value, if so desired (e.g., $H_M$: $M_1 - M_2 = 0 \pm$ MES, $\alpha$ = .05, $CV_t = 2.38$).

---

*Caution: Neyman-Pearson's critical region is very similar to the equivalent critical region you would obtain by using Fisher's sig as a cut-off point on a null distribution. However, this similarity is rather unimportant on three accounts: it is based on a critical value which delimits the region to reject $H_M$ in favor of $H_A$, irrespective of the actual observed value of the test (Fisher, on the contrary, is more interested in the actual p-value of the research result); it is fixed a priori and, thus, rigid and immobile (Fisher's level of significance can be flexible—Macdonald, 2002); and it is non-gradable (with Fisher's approach, you may delimit several more extreme critical regions as areas of stronger evidence).*

---

**Step 4 - Set up the alternative hypothesis ($H_A$).** One of the main innovations of Neyman-Pearson's approach was the consideration of alternative hypotheses (Neyman, 1956; Neyman and Pearson, 1928, 1933). Unfortunately, the alternative hypothesis is often postulated in an unspecified manner (e.g., as $H_A$: $M_1 - M_2 \neq 0$), even by Neyman and Pearson themselves (Jones and Tukey, 2000; Macdonald, 1997). In practice, a fully specified alternative hypothesis (e.g., its mean and variance) is not necessary because this hypothesis only provides partial information to the testing of the main hypothesis (a.k.a., the effect size and β). Therefore, the alternative hypothesis is better written so as for incorporating the minimum effect size within its postulate (e.g., $H_A$: $M_1 - M_2 \neq 0 \pm$ MES). This way it is clear that values beyond such minimum effect size are the ones considered of research importance.

---

*Caution: Neyman-Pearson's $H_A$ is often postulated as the negation of a nil hypothesis ($H_A$: $M_1 - M_2 \neq 0$), which is coherent with a simple postulate of $H_M$ ($H_M$: $M_1 - M_2 = 0$). These simplified postulates are not accurate and are easily confused with Fisher's approach to data testing—$H_M$ resembles Fisher's $H_0$, and $H_A$ resembles a mere negation of $H_0$. However, merely negating $H_0$ does not make its negation a valid alternative hypothesis—otherwise Fisher would have put forward such alternative hypothesis, something which he was vehemently against (Hubbard, 2004). As discussed earlier, Neyman-Pearson's approach introduces the construct of effect size into their testing approach; thus, incorporating such construct in the specification of both $H_M$ and $H_A$ makes them more accurate, and less confusing, than their simplified versions.*

---

Among things to consider when setting the alternative hypothesis are the expected effect size in the population (see above) and the Type II error you are prepared to commit.

**Type II error.** A Type II error (or error of the second class) is made every time the main hypothesis is wrongly retained (thus, every time $H_A$ is wrongly rejected). Making a Type II error is less critical than making a Type I error, yet you still want to minimize the probability of making this error once you have decided which alpha level to use (Macdonald, 2002; Neyman, 1942; Neyman and Pearson, 1933).

**Beta (β).** Beta is the probability of committing a Type II error in the long run and is, therefore, a parameter of the alternative hypothesis (Figure 4, Neyman, 1956). You want to make beta as small as possible, although not smaller than alpha (if β needed to be smaller than α, then $H_A$ should be your main hypothesis, instead!). Neyman and Pearson proposed 20% (β = .20) as an upper ceiling for beta, and the value of alpha (β = α) as its lower floor (Neyman, 1953). For symmetry with the main hypothesis, the alternative hypothesis can, thus, be written so as for incorporating the beta level in its postulate (e.g., $H_A$: $M_1 - M_2 \neq 0 \pm$ MES, β = .20).

**Step 5 - Calculate the sample size (N) required for good power (1-β).** Neyman-Pearson's approach is eminently a priori in order to ensure that the research to be done has good power (Macdonald, 2002; Neyman, 1942, 1956; Pearson, 1955). Power is the probability of correctly rejecting the main hypothesis in favor of the alternative hypothesis (i.e., of correctly accepting $H_A$). It is the mathematical opposite of the Type II error (thus, $1 - β$; Hubbard, 2004; Macdonald, 1997). Power depends on the type of test selected (e.g., parametric tests and one-tailed tests increase power), as well as on the expected effect size (larger ES's increase power), alpha (larger α's increase power) and beta (smaller β's increase power). A priori power is ensured by calculating the correct sample size given those parameters (Spielman, 1973). Because power is the opposite of beta, the lower floor for good power is, thus, 80% ($1 - β = .80$), and its upper ceiling is alpha ($1 - β = α$).

---

*Note: $H_A$ does not need to be tested under Neyman-Pearson's approach, only $H_M$ (Neyman, 1942; Neyman and Pearson, 1928, 1933; Pearson, 1955; Spielman, 1973). Therefore, the procedure looks similar to Fisher's and, under similar circumstances (e.g., when using the same test and sample size), it will lead to the same results. The main difference between procedures is that Neyman-Pearson's $H_A$ provides explicit information to the test; that is, information about ES and β. If this information is not taken into account for designing a research project with adequate power, then, by default, you are carrying out a test under Fisher's approach.*

*Caution: For Neyman and Pearson, there is little justification in carrying out research projects with low power. When a research project has low power, Type II errors are too big, so it is less probable to reject $H_M$ in favor of $H_A$, while, at the same time, it makes unreasonable to accept $H_M$ as the best explanation for the research results. If you face a research project with low a priori power, try the best compromise between its parameters (such as increasing α, relaxing β, settling for a larger ES, or using one-tailed tests; Neyman and Pearson, 1932). If all fails, consider Fisher's approach, instead.*

---

**Step 6 - Calculate the critical value of the test ($CV_{test}$, or $Test_{crit}$).** Some of above parameters (test, α and N) can be used for calculating the critical value of the test; that is, the value to be used as the cut-off point for deciding between hypotheses (Figure 5, Neyman and Pearson, 1933).

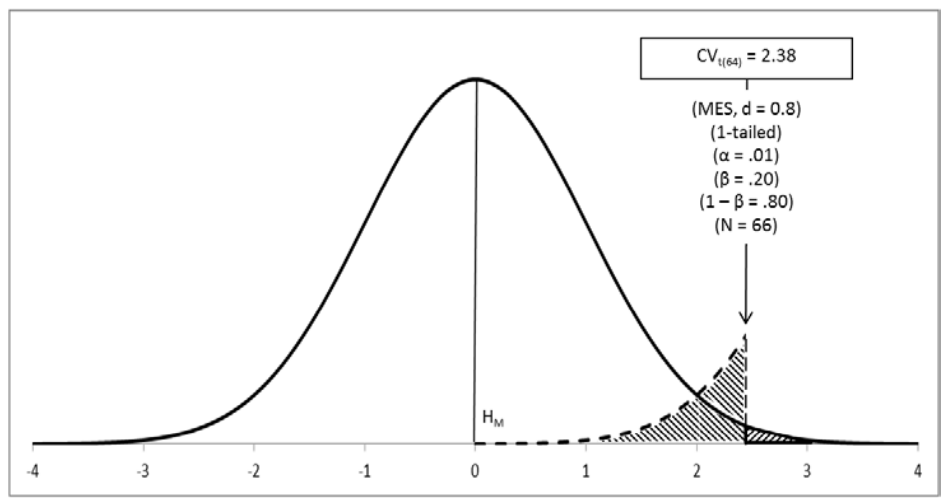

*Figure 5.* Neyman-Pearson's test in action: $CV_{test}$ is the point for deciding between hypotheses; it coincides with the cut-off points underlying α, β and MES.

## 3.2.    A posteriori steps

**Step 7 - Calculate the test value for the research (RV$_{test}$).** In order to carry out the test, some unknown parameters of the populations are estimated from the sample (e.g., variance), while other parameters are deduced theoretically (e.g., the distribution of frequencies under a particular statistical distribution). The statistical distribution so established thus represents the random variability that is theoretically expected for a statistical main hypothesis given a particular research sample, and provides information about the values expected at different locations under such distribution.

By applying the corresponding formula, the research value of the test (RV$_{test}$) is obtained. This value is closer to zero the closer the research data is to the mean of the main hypothesis; it gets larger the further away the research data is from the mean of the main hypothesis.

---

*Note:* P-*values can also be used for testing data when using Neyman-Pearson's approach, as testing data under H$_M$ is similar to testing data under Fisher's H$_0$ (Fisher, 1955). It implies calculating the probability of the research data under the distribution of H$_M$—P(D|H$_M$). Just be mindful that* p-*values go in the opposite way than RVs, with larger* p-*values being closer to H$_M$ and smaller* p-*values being further away from it.*

*Caution: Because of above equivalence, you may use* p-*values instead of CV$_{test}$ with Neyman-Pearson's approach. However,* p-*values need to be considered mere proxies under this approach and, thus, have no evidential properties whatsoever (Frick, 1996; Gigerenzer, 2004). For example, if working with a priori α = .05, p = .01 would lead you to reject H$_M$ at α = .05; however, it would be incorrect to reject it at α = .01 (i.e., α cannot be adjusted a posteriori), and it would be incorrect to conclude that you reject H$_M$ strongly (i.e., α cannot be gradated). If confused, you are better off sticking to CV$_{test}$, and using* p-*values only with Fisher's approach.*

---

**Step 8 - Decide in favor of either the main or the alternative hypothesis.** Neyman-Pearson's approach is rather mechanical once the a priori steps have been satisfied (Macdonald, 2002; Neyman, 1942, 1956; Neyman and Pearson, 1933; Spielman, 1978). Thus, the analysis is carried out as per the optimal test selected and the interpretation of results is informed by the mathematics of the test, following on the a priori pattern set up for deciding between hypotheses:

- If the observed result falls within the critical region, reject the main hypothesis and accept the alternative hypothesis.

- If the observed result falls outside the critical region and the test has good power, accept the main hypothesis.

- If the observed result falls outside the critical region and the test has low power, conclude nothing. (Ideally, you would not carry out research with low power—Neyman, 1955).

---

*Notes: Neyman-Pearson's approach leads to a decision between hypotheses (Neyman and Pearson, 1933; Spielman, 1978). In principle, this decision should be between rejecting H$_M$ or retaining H$_M$ (assuming good power), as the test is carried out on H$_M$ only (Neyman, 1942). In practice, it does not really make much difference whether you accept H$_M$ or H$_A$, as appropriate (Macdonald, 1997). In*

*fact, accepting either $H_M$ or $H_A$ is beneficial as it prevents confusion with Fisher's approach, which can only reject $H_0$ (Perezgonzalez, 2014).*

*Reporting the observed research test value is relevant under Neyman-Pearson's approach, as it serves to compare the observed value against the a priori critical value—e.g., t(64) = 3.31, 1-tailed > $CV_t$ = 2.38, thus accept $H_A$. When using a p-value as a proxy for $CV_{test}$, simply strip any evidential value off p—e.g., t(64) = 3.31, p < α, 1-tailed.*

*Neyman-Pearson's hypotheses are also assumed to be true. $H_M$ represents the probability distribution of the data given a true hypothesis—$P(D/H_M)$, while $H_A$ represents the distribution of the data under an alternative true hypothesis—$P(D/H_A)$, even when it is never tested. This means that $H_M$ and $H_A$ cannot be, at the same time false, nor proved or falsified a posteriori. The only way forward is to act as if the conclusion reached by the test was true—subject to a probability α or β of making a Type I or Type II error, respectively (Cortina and Dunlap, 1997; Neyman and Pearson, 1933).*

### 3.3. Highlights of Neyman-Pearson's approach

**More powerful.** Neyman-Pearson's approach is more powerful than Fisher's for testing data in the long run (Williams, Zimmerman, Ross, and Zumbo 2006). However, repeated sampling is rare in research (Fisher, 1955).

**Better suited for repeated sampling projects.** Because of above, Neyman-Pearson's approach is well-suited for repeated sampling research using the same population and tests, such as industrial quality control or large scale diagnostic testing (Fisher, 1955; Spielman, 1973).

**Deductive.** The approach is deductive and rather mechanical once the a priori steps have been set up (Fisher, 1955; Neyman, 1942; Neyman and Pearson, 1933).

**Less flexible than Fisher's approach.** Because most of the work is done a priori, this approach is less flexible for accommodating tests not thought of beforehand and for doing exploratory research (Macdonald, 2002).

**Defaults easily to Fisher's approach.** As this approach looks superficially similar to Fisher's, it is easy to confuse both and forget what makes Neyman-Pearson's approach unique (Lehman, 1993). If the information provided by the alternative hypothesis—ES and β—is not taken into account for designing research with good power, data analysis defaults to Fisher's test of significance.

### 4. Null hypothesis significance testing

NHST is the most common procedure used for testing data nowadays, albeit under the false assumption of testing substantive hypotheses (Carver, 1978; Hager, 2013; Hubbard, 2004; Nickerson, 2000). NHST is, in reality, an amalgamation of Fisher's and Neyman-Pearson's theories, offered as a seamless approach to testing (Gigerenzer, 2004; Macdonald, 2002). It is not a clearly defined amalgamation either and, depending on the author describing it or on the researcher using it, it may veer more towards Fisher's approach (e.g., American Psychological Association, 2010; Krueger, 2001; Nunnally, 1960; Wilkinson and the Task Force on Statistical Inference, 1999) or towards Neyman-Pearson's approach (e.g., Cohen, 1988; Cortina and Dunlap, 1997; Frick, 1996; Kline, 2004; Nickerson, 2000; Rosnow and Rosenthal, 1989; Schmidt, 1996; Wainer, 1999).

Unfortunately, if we compare Fisher's and Neyman-Pearson's approaches vis-à-vis, we find that they are incompatible in most accounts (Table 1). Overall, however, most amalgamations follow Neyman-Pearson procedurally but Fisher philosophically (Cortina and Dunlap, 1997; Hubbard, 2004; Johnstone, 1986; Spielman, 1978).

**Table 1 |** Equivalence of constructs in Fisher's and Neyman-Pearson's theories, and amalgamation of constructs under NHST.

| Concept | | Fisher | | Neyman-Pearson |
|---|---|---|---|---|
| **Test object** | | Data—$P(D|H_0)$ | = | Data—$P(D|H_M)$ |
| | NHST | | ➦ Data as if testing a falsifiable hypothesis—$P(H_0|D)$ | ➥ |
| **Approach** | | A posteriori | ≠ | A priori |
| | NHST | | ➦ A posteriori, sometimes both | ➥ (partly) |
| **Research goal** | | Statistical significance of research results | ≠ | Deciding between competing hypotheses |
| | NHST | | ➦ Statistical significance, also used for deciding between hypotheses | ➥ |
| **Hs under test** | | $H_0$, to be nullified with evidence | ≈ | $H_M$, to be favored against $H_A$ |
| | NHST | | ➦ Both ($H_0 = H_M$) | ➥ |
| **Alternative hypothesis** | | Not needed (implicitly, 'No $H_0$') | ≠ | Needed. Provides ES and β |
| | NHST | | ➦ $H_A$ posed as 'No $H_0$' (ES and β sometimes considered) | ➥ (partly) |
| **Prob. distr. of test** | | As appropriate for $H_0$ | = | As appropriate for $H_M$ |
| | NHST | | ➦ As appropriate for $H_0$ | ➥ |
| **Cut-off point** | | Sig identifies noteworthy results; can be gradated; can be corrected a posteriori | ≠ | Common to $CV_{test}$, α, β, and MES; cannot be gradated; cannot be corrected a posteriori |
| | NHST | | ➦ Sig = α, can be gradated, can be corrected a posteriori | ➥ (partly) |
| **Sample size calculator** | | None | ≠ | Based on test, ES, α, and power (1- β) |
| | NHST | | ➦ Either | ➥ |
| **Statistic of interest** | | *p*-value, as evidence against $H_0$ | ≠ | $CV_{test}$ (*p*-value has no inherent meaning but can be used as a proxy instead) |
| | NHST | | ➦ *p*-value, used both as evidence against $H_0$ and a proxy to accept $H_A$ | ➥ |
| **Error prob.** | | α possible, but irrelevant with single studies (partly) ➦ | ≠ | α = Type I error prob. β = Type II error prob. |
| | NHST | | *p*-value = α = Type I error in single studies (β sometimes considered) | ➥ (partly) |

| | | | |
|---|---|---|---|
| **Result falls outside critical region**<br><br>NHST | Ignore result as not significant<br>↳ | $\neq$<br><br>Either ignore result as not significant; or accept $H_0$; or conclude nothing | Accept $H_M$ if good power; conclude nothing otherwise<br>↳ |
| **Result falls in critical region**<br><br>NHST | Reject $H_0$<br>↳ | $\neq$<br><br>Either | Accept $H_A$ (= Reject $H_M$ in favor of $H_A$)<br>↳ |
| **Interpretation of results in critical region**<br><br>NHST | Either a rare event occurred or $H_0$ does not explain the research data | $\neq$<br><br>$H_A$ has been proved / is true; or $H_0$ has been disproved / is false; or both | $H_A$ explains research data better than $H_M$ does (given $\alpha$) |
| **Next steps**<br><br>NHST | Rejecting $H_0$ does not automatically justify not $H_0$. Replication needed, meta-analysis is useful. | $\neq$<br><br>None (results taken as definitive, especially if significant); further studies may be sometimes recommended (especially if results are not significant) | Impossible to know whether $\alpha$ error has been made. Repeated sampling of same population needed, Monte Carlo is useful. |

NHST is not only ubiquitous but very well ingrained in the minds and current practice of most researchers, journal editors and publishers (Hubbard, 2004; Gigerenzer, 2004; Spielman, 1978), especially in the biological sciences (Lovel, 2013; Ludbrook, 2013), social sciences (Frick, 1996), psychology (Gigerenzer, 2004; Nickerson, 2000) and education (Carver, 1978, 1993). Indeed, most statistics textbooks for those disciplines still teach NHST rather than the two approaches of Fisher and of Neyman and Pearson as separate and rather incompatible theories (e.g., Dancey and Reidy, 2014). NHST has also the (false) allure of being presented as a procedure for testing substantive hypotheses (Gigerenzer, 2004; Macdonald, 2002).

In the situations in which they are most often used by researchers, and assuming the corresponding parameters are also the same, both Fisher's and Neyman-Pearson's theories work with the same statistical tools and produce the same statistical results; therefore, by extension, NHST also works with the same statistical tools and produces the same results—in practice, however, both approaches start from different starting points and lead to different outcomes (Berger, 2003; Fisher, 1955; Spielman, 1978). In a nutshell, the differences between Fisher's and Neyman-Pearson's theories are mostly about research philosophy and about how to interpret results (Fisher, 1955).

The most coherent plan of action is, of course, to follow the theory which is most appropriate for purpose, be this Fisher's or Neyman-Pearson's. It is also possible to use both for achieving different goals within the same research project (e.g., Neyman-Pearson's for tests thought of a priori, and Fisher's for exploring the data further, a posteriori), pending that those goals are not mixed up.

However, the apparent parsimony of NHST and its power to withstand threats to its predominance are also understandable. Thus, I propose two practical solutions to improve NHST: the first a compromise to improve Fisher-leaning NHST, the second a compromise to improve Neyman-Pearson-leaning NHST. A computer program such as G*Power can be used for implementing the recommendations made for both.

### 4.1.    Improving Fisher-leaning NHST

Fisher's is the closest approach to NHST; it is also the philosophy underlying common statistics packages, such as SPSS. Furthermore, because using Neyman-Pearson's concepts within NHST may be irrelevant or inelegant but hardly damaging, it requires little re-engineering. A clear improvement to NHST comes from incorporating Neyman-Pearson's constructs of effect size and of a priori sample estimation for adequate power. Estimating effect sizes (both a priori and a posteriori) ensures that researchers consider importance over mere statistical significance. A priori estimation of sample size for good power also ensures that the research has enough sensitiveness for capturing the expected effect size (Huberty, 1987; Macdonald, 2002).

### 4.2.    Improving Neyman-Pearson-leaning NHST

NHST is particularly damaging for Neyman-Pearson's approach, simply because the later defaults to Fisher's if important constructs are not used correctly. An importantly damaging issue is the assimilation of *p*-values as evidence of Type I errors and the subsequent correction of alphas to match such *p*-values (roving α's, Goodman, 1993; Hubbard, 2004). The best compromise for improving NHST under these circumstances is to compensate a posteriori roving alphas with a posteriori roving betas (or, if so preferred, with a posteriori roving power). Basically, if you are adjusting alpha a posteriori (roving α) to reflect both the strength of evidence (sig) and the long-run Type I error (α), you should also adjust the long-run probability of making a Type II error (roving β). Report both roving alphas and roving betas for each test, and take them into account when interpreting your research results.

> **Caution:** NHST is very controversial, even if the controversy is not well known. A sample of helpful readings on this controversy are Christensen (2005), Hubbard (2004), Gigerenzer (2004), Goodman (1999), Louçã (2008, https://www.repository.utl.pt/bitstream/10400.5/2327/1/wp022008.pdf), Halpin and Stam (2006), Huberty (1993), Johnstone (1986), and Orlitzky (2012).

### 5.    Conclusion

Data testing procedures represented a historical advancement for the so-called "softer" sciences, starting in biology but quickly spreading to psychology, the social sciences and education. These disciplines benefited from the principles of experimental design, the rejection of subjective probabilities and the application of statistics to small samples that Sir Ronald Fisher started popularizing in 1922 (Lehmann, 2011), under the umbrella of his tests of significance (e.g., Fisher, 1954). Two mathematical contemporaries, Jerzy Neyman and Egon Sharpe Pearson, attempted to improve Fisher's procedure and ended up developing a new theory, one for deciding between competing hypotheses (Neyman and Pearson, 1928), more suitable to quality control and large scale diagnostic testing (Spielman, 1973). Both theories had enough similarities to be easily confused (Perezgonzalez, 2014), especially by those less epistemologically inclined; a confusion fiercely opposed by the original authors (e.g., Fisher, 1955)—and ever since (e.g., Hager, 2013; Lehmann, 2011; Nickerson, 2000)—but something that irreversibly happened under the label of null hypothesis significance testing. NHST is an incompatible amalgamation of the theories of Fisher and of Neyman and Pearson (Gigerenzer, 2004). Curiously, it is an amalgamation that is technically reassuring despite it being, philosophically, pseudoscience. More interestingly, the numerous critiques raised against it for the past 80 years have not only failed to debunk NHST from the researcher's statistical toolbox, they have also failed to be widely known, to find their way into statistics manuals, to be edited out of journal submission requirements, and to be flagged up by peer-reviewers (e.g., Gigerenzer, 2004). NHST effectively negates the benefits that could be gained from Fisher's and from Neyman-Pearson's theories; it also slows scientific progress (Savage, 1957; Carver, 1978, 1993) and may be fostering pseudoscience. The best option would be to ditch NHST altogether and revert to the theories of Fisher and of Neyman-Pearson as—and when—appropriate. For everything else, there are alternative tools, among them exploratory data analysis (Tukey, 1977), effect sizes (Cohen, 1988), confidence intervals (Neyman, 1935), meta-analysis (Rosenthal, 1984), Bayesian applications (Dienes, 2014) and, chiefly, honest critical thinking (Fisher, 1960).